\documentclass{easychair}
\usepackage{amsmath}
\usepackage[latin1]{inputenc}
\usepackage[english]{babel}
\usepackage[T1]{fontenc}
\usepackage{graphicx}
\usepackage{amsfonts}
\usepackage{float}
\usepackage{graphics}

\newcommand{\oen}{[\hspace*{-4.80pt}\ [} 
\newcommand{\cen}{]\hspace*{-4.80pt}\ ]} 
\title{A Model for Interactive Scores with Temporal Constraints and Conditional Branching} 

\author{
Mauricio Toro\inst{1}
\and
Myriam Desainte-Catherine\inst{2}
\and
Pascal Baltazar\inst{3}
}

\institute{
  Universidad Eafit,
  Medellin, Colombia\\
\and
   LABRI,
   Bordeaux, France\\
\and
   GMEA,
   Albi, France\\
 }

\begin{document}
\maketitle
\begin{abstract}
Interactive Scores (IS) are a formalism for the design and performance of interactive multimedia scenarios. 
IS provide temporal relations (TR), but they cannot represent conditional branching and TRs simultaneously.
We propose an extension to Allombert et al.'s IS model by including a condition on the TRs.
We found out that in order to have a coherent model in all possible scenarios, durations must be flexible; however, sometimes it is possible to have fixed durations.
To show the relevance of our model, we modeled an existing multimedia installation called Mariona.
In Mariona there is choice, 
random durations and loops. 
Whether we can represent all the TRs available in Allombert et al.'s model into ours, or we have to
choose between a timed conditional branching model and a pure temporal model before writing
a scenario, still remains as an open question.

\end{abstract}
\section{Introduction}\label{sec:introduction}

\textit{Interactive Scores (IS)} are a formalism for the design and performance of scenarios represented by \textit{temporal objects (TO)}, \textit{temporal relations (TR)} and discrete interactive events. Examples of TOs are videos, sounds or audio processors. TOs can be triggered by interactive events (usually launched by the user) and several TOs can be active simultaneously. TRs are used, for instance, to express precedence between objects, relations on their durations and explicit values for their durations.



IS have been subject of study since the beginning of the century \cite{bdc01}, \cite{bdc03}. IS were originally developed for interactive music scores.
Recently, the model was extended by Allombert, Desainte-Catherine, Laralde and Assayag in \cite{artech2008}.
Hence IS can describe any kind of TOs, Allombert \textit{et al.}'s model has inspired two applications: \textit{iScore} \cite{aadc08} to compose and perform Electroacoustic music and \textit{Virage} \cite{bamcrs09} to control image, video, audio and lights on live spectacles and interactive museums.

Allombert \textit{et al.} showed on \textit{iScore} and \textit{Virage} that IS are successful to describe TRs, but IS have not been used to represent scenarios that require \textit{conditional branching}.
Conditional branching is commonly used in programming to describe control structures (e.g., \textit{if/else} and \textit{switch/case}).
It provides a mechanism to choose the state of a program based on a condition and its current state.

As in programming, using conditional branching, a designer can create scenarios with loops, concurrent
execution of multiple instances of the same TO and choice.
Using conditional branching, the user or the system can take decisions on the performance of the scenario 
with the degree of freedom that the designer described --while the system maintaints the TRs of the scenario.
For instance, the designer can specify a condition to end a loop: ``when the user change the value of the variable $end$ to \texttt{true}, the loop stops''. The designer can also specify that such choice is made by the system: ``the system non-deterministically chooses to stop the loop or to continue''.

Allombert \textit{et al.} represent conditional branching
and TRs separately, but there is not an unified way to represent conditional branching together with \\quantitative and qualitative TRs in the same scenario. 

\textit{Quantitative TRs} are those involving a proportional or explicit duration; for instance, ``the duration of $A$ is one third of the duration of $B$'' or ``the duration of $A$ is 3 seconds''. On the other hand, \textit{qualitative TRs} represent precedence between the start and end points of two TOs; for instance, ``$A$ must be played during $B$'' or ``$C$ must be played after $D$''.

\textbf{In this paper we propose a new model for IS. It extends Allombert \textit{et al.}'s model by including a condition on the qualitative TRs.}
We do not include quantitative TRs because we found out that durations must be \textit{flexible} (i.e., they can have any duration) to be coherent in all scenarios. However, we show that in some scenarios 
it is possible to respect \textit{rigid} durations (i.e., durations with values in a finite interval).

The remainder of this paper is structured as follows. Section 2 shows related work on formalisms and applications for interactive multimedia. Section 3 presents our model for Interactive Scores. Section 4 explains the relation between our model and Allombert \textit{et al.}'s model. Section 5 shows how to represent fragments of \textit{Mariona}\footnote{\url{http://www.gmea.net/activite/creation/2007\_2008/pPerez.htm} } in our model. Finally, section 6 gives concluding remarks and proposes future works.

\section{Related Work}

\textit{Mariona} is a multimedia installation, created by Pol Perez in 2007, that includes temporal relations, conditional branching, random durations, choice and hierarchy (i.e., an object can contain other objects). Mariona has a vision-based hand and body tracking system, speakers and a video display. A similar installation that uses motion sensors, instead of a tracking system, is described in \cite{YI07}.
Both installations are controlled by a Max/MSP \cite{max} program --like most interactive multimedia applications. 
\subsection{Applications for Interactive Multimedia}
In the domain of interactive music, there are applications such as \textit{Ableton Live}\footnote{\url{http://www.ableton.com/live/} }. Using \textit{Live}, the composer can write loops and the musician can control
different parameters of the piece during performance.

An application to define a hierarchy and temporal relations among temporal objects is OpenMusic Maquettes \cite{Bresson05b}. Unfortunately,
OpenMusic is designed for composition and not real-time interaction. 

Another model related to Interactive Scores (IS) is \textit{score following} \cite{cont08a}.
Such systems ``follow'' the performance a real instrument and may play multimedia associated to certain notes in the score of the piece.
However, to use these systems it is necessary to play a real instrument. On the other hand, using IS the user only has to control some parameters of the piece such as the date of the events, and the system plays the temporal objects described on the score.

\subsection{Formalisms for Interactive Multimedia}
To handle more complex synchronization patterns and to predict the behavior of interactive scenarios, 
formalisms such as the \textit{Hierarchical Time Stream Petri Networks (HTSPN)} \cite{SSW95} and the \textit{Non-determi- nistic Timed Concurrent
Constraint Programming} (\texttt{ntcc}) process calculus \cite{ntcc} have been used to model IS in \cite{artech2008} and \cite{AADR06}, respectively.
Formalisms based on process calcus have been used for the modeling of interactive music systems 
 \cite{is-chapter,tdcr14,ntccrt,cc-chapter,torophd,torobsc,Toro-Bermudez10,Toro15,ArandaAOPRTV09,tdcc12,toro-report09,tdc10,tdcb10,tororeport} 
 and the modeling and analysis ecological systems \cite{PT13, TPSK14, PTA13, mean-field-techreport}. 

In HTSPN we can to express a variety of temporal relations, but it is not easy to represent global constraints (e.g.,
 the number of temporal objects playing simultaneously). On the other hand, \texttt{ntcc} makes it possible to synchronize processes through a common constraint
\textit{store}, thus global constraints are explicitly represented in such store. We plan to write our model on \texttt{ntcc} because we can easily represent time, constraints, choice, and we can verify the model.

An advantage of using formal methods to model interactive multimedia is that they usually have automatic verification techniques. There are numerous studies to verify liveness, fairness, reachability and boundness on Petri Networks \cite{tm88}. On the other hand, in the last years, calculi similar to \texttt{ntcc} have been subject of study for automatic model checking procedures \cite{fv06}.

For instance, using formal methods, it is possible to verify that will not be deadlocks, and also that certain temporal objects will be reached (played) during performance. This kind of properties are not possible to verify on applications for interactive multimedia with no formal semantics.

\section{Interactive Scores with Conditional Branching}
In this section we show how we can extend Allombert \textit{et al.}'s model with conditional branching. 
We also show the new possibilities that our model offers.

Our model is based on the concept of \textit{points}, and  
 we provide relations for the points. The \textit{before} relation 
is the only type of relation in our model. A relation $p$ \textit{before} $q$ means that
the execution of $q$ is preceded by the execution of $p$ if the condition in the relation holds. 
Although relations also
have a nominal duration, it may change during the performance. The nominal duration is computed during the edition of the scenario using constraint programming.

Relations and temporal objects build up Interactive Scores (IS), thus  
a \textit{score}\footnote{We still use the term \textit{score} for historical reasons.} (i.e., the specification of a scenario) is defined by a tuple $s = \langle T,R \rangle$, where $T$ is a set of temporal objects and $R$ is a set of relations.  Relations and temporal objects are described using points.

\subsection{Points}
A \textit{Point} is defined by $p = \langle D, b_p, b_s \rangle$, where $D \subseteq \mathbb{N}$ is the set of its possible dates of execution.
Intuitively, we say that point $p$ is a \textit{precedessor} of $q$, if there is a relation $p$ \textit{before} $q$. On the other hand,
we say that a point $p$ is a \textit{sucessor} of $r$, if there is a relation $r$ \textit{before} $p$.

There are two behaviors for a point. $b_p$ defines
whether the point waits until all its predecessors transfer the control to it --\textit{Wait for All (WA)}-- or it only waits for the first of them --\textit{Wait for the First (WF)}--. $b_s$ defines whether the point transfers the control to all its successors which conditions hold --\textit{No CHoice (NCH)}--  or it chooses one of them --\textit{CHoice (CH)}--.
In this model, a point can only belong to a single temporal object to avoid ambiguities.

\subsection{Temporal Objects}
A \textit{temporal object (TO)} is defined by $t = \langle p_s, p_e, c,\\ d, proc, param, N , vars \rangle$, where $p_s$ is a point that starts a new instance of $t$ and $p_e$ ends such instance; 
$c$ is a constraint attached to $t$ (i.e., a constraint with local information for $t$ and its children); $d$ is the duration; $proc$ is a process which executes along with $t$;  $param$ are the parameters for the process; 
and $N$ is the set of TOs embedded in $t$, which are called children of $t$. 
Finally, $vars$ represents the local variables defined for the TO. Local variables can be used by $t$'s childrens,  process and local constraint.

The reader may notice that a TO does not provide a relation between its start and end points.
For that reason, we must define a Timed Conditional Relation (TCR) between the start and the end points of the TO.
We must also define a TCR between the start point of a father and the start point of at least one of its children.

\subsection{Timed Conditional Relations}
A \textit{Timed Conditional Relation (TCR)} is defined by $r = \langle p_1, p_2, c, d, b, e\rangle$, where $p_1$ and $p_2$ are the points involved in the relation, $c$ is the condition to determine whether the control \textit{jumps} from $p_1$ to $p_2$ (i.e., the control is transferred from $p_1$ to $p_2$), $d$ is the duration of the relation,  $b$ is the interpretation for $c$,  $e$ describes whether the condition is evaluated as soon as it holds, or at the end of the duration of the relation. 

In what follows we explain some of the parameters of a TCR. 
We recall from \cite{artech2008} that a duration is \textit{flexible} if it can take any value, \textit{rigid} if it takes values between two fixed integers and \textit{semi-rigid} if it takes values greater than a fixed integer. In our model, we only guarantee the coherence of flexible durations. When the durations are flexible, there are two possible values for $e$: \textit{now} or \textit{wait}; otherwise, only \textit{wait} is possible. Finally, there are two possible values for $b$: 
\textit{when} means that if $c$ holds, the control jumps; \textit{unless} means that if $c$ does not hold or its value cannot be deduced from the environment (e.g., $c = a > 0$ and $-\infty < a < \infty$), the control jumps.


\subsection{Example: A Score with a Conditional Loop}

The following example describes a score with a loop. During the execution, the system plays the sound $B$, a silence of one second, and
the video $C$. If the signal $finish$ becomes true, it ends the scenario after playing the video $C$; otherwise, it
comes back to the beginning of the sound $B$ after playing the video $C$. To define the score of this scenario, we define a local boolean variable
$finish$ inside the structure $A$ and we use it as the condition for the relations. 

\begin{figure}[h!]
  \begin{center}
    \mbox{
     
{\includegraphics[width=\columnwidth]{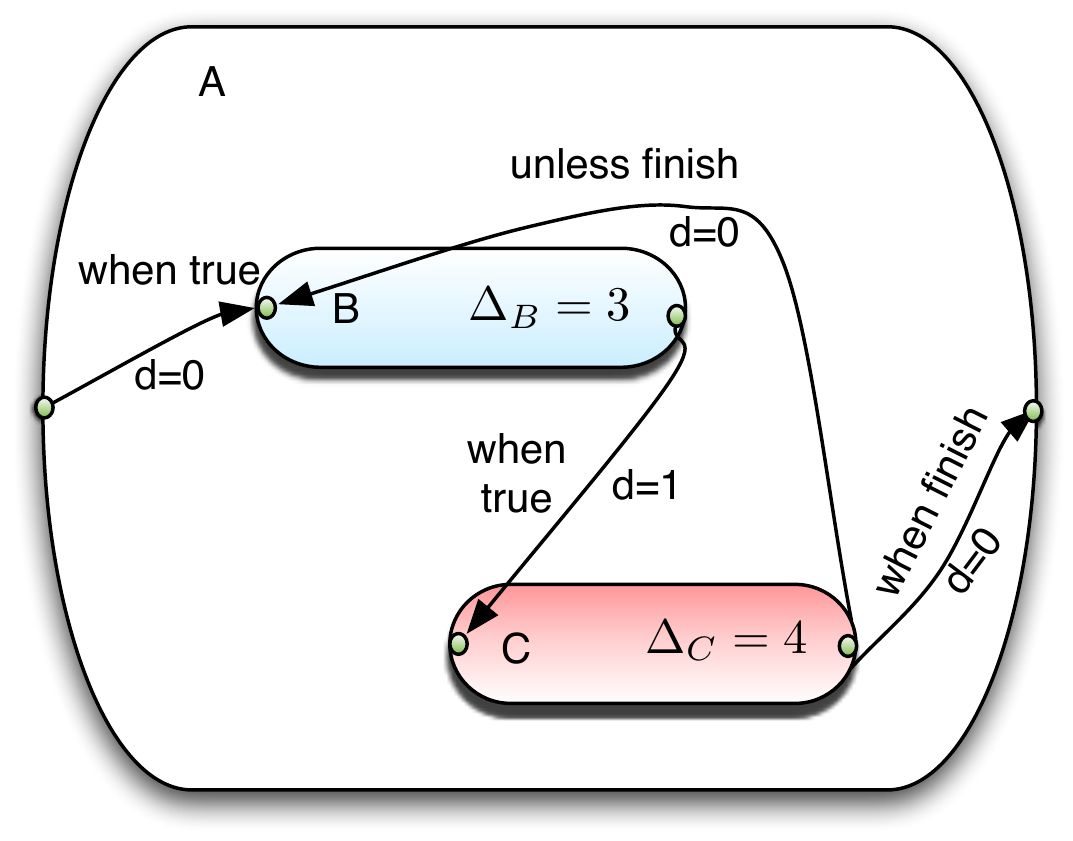}}  
 
      }
    \caption{A score with conditional branching.}
    \label{fig:basic-example}
  \end{center}
\end{figure}

Figure \ref{fig:basic-example} is a representation of the scenario. The duration for $B$ is three seconds and the duration for $C$ is four seconds (these values are calculated during the edition phase); 
however, they can change during the performance of the scenario depending on the behavior of the points and the relations.

The points have the following behavior. Point $e_c$ (the end of the TO $C$) is enabled for choice and the other points transfer the control to all their successors. All the points wait for the first
predecessor that transfer the control to them. Finally, the relations wait until their durations finish before evaluating their conditions, preserving rigid durations. Formally, \\

\noindent
$
s_A = s_B = \langle \{d,d \geq 0 \land d \% 8 = 0\}, WF, NCH \rangle\\
s_C = \langle \{d, d \geq 4 \land (d - 4)\% 8 = 0\}, WF, NCH \rangle\\
e_A = \langle\{d, d \geq 8 \land( d - 8) \% 8 = 0\}, WF, NCH \rangle \\
e_B = \langle \{d, d \geq 3 \land (d - 3) \% 8 = 0\}, WF,NCH \rangle\\
e_C = \langle \{d, d \geq 8 \land (d - 8) \% 8 = 0\}, WF, CH \rangle\\
B = \langle s_B, e_B, \texttt{true}, 3, playSoundB, \phi, \phi, \phi \rangle \\
C = \langle s_C, e_C, \texttt{true}, 4, playVideoC, \phi, \phi, \phi \rangle\\
A = \langle s_A, e_A, \texttt{true}, 8, silence\footnote{$silence$ is a process that does not perform any action.}, \phi, \{ B, C \}, \\
\hspace*{23pt} \{finish\} \rangle\\
T = \{A \}\\
R = \{ \langle s_A, s_B, \texttt{true}, 0, when, wait\rangle, \\
\hspace*{23pt} \langle e_B, s_C, \texttt{true}, 1, when, wait \rangle, \\
\hspace*{23pt} \langle e_C, e_A, finish, 0, when, wait \rangle, \\
\hspace*{23pt} \langle e_C, s_B, finish, 0, unless, wait \rangle,\\
\hspace*{23pt} \langle s_B, e_B, \texttt{true}, 3, when, wait \rangle,\\
\hspace*{23pt} \langle s_C, e_C, \texttt{true}, 4, when, wait \rangle\}\\
s = \{T,R\}
$

\subsection{Multiple Instances Concurrently}
Some scenarios require multiple instances of the same TO executing concurrently. This must be treated in a special way because
not all the processes associated to a TO accept that sort of \textit{polyphony}. Consider, for instance, a TO that displays a video.  There may not be a defined behavior for more than one instance of it 
at the same time.

\begin{figure}[h!]
  \begin{center}
    \mbox{
     
{\includegraphics[width=\columnwidth]{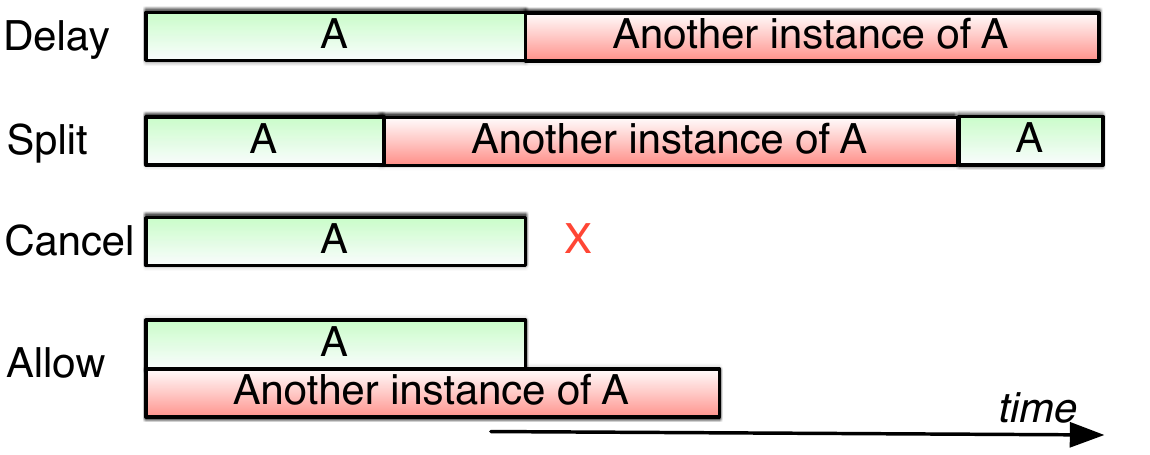}}  
 
      }
    \caption{Multiple instances concurrently.}
    \label{fig:concurrentinstances}
  \end{center}
\end{figure}

We propose four behaviors (fig. \ref{fig:concurrentinstances}) to manage concurrent instances of the same TO: splitting them, delaying them,
cancelling them or allowing them.
This behavior must be selected by the designer of the scenario according to the nature of the TO. 

\subsection{Limitations}

\begin{figure}[h!]
  \begin{center}
    \mbox{
     
{\includegraphics[width=\columnwidth]{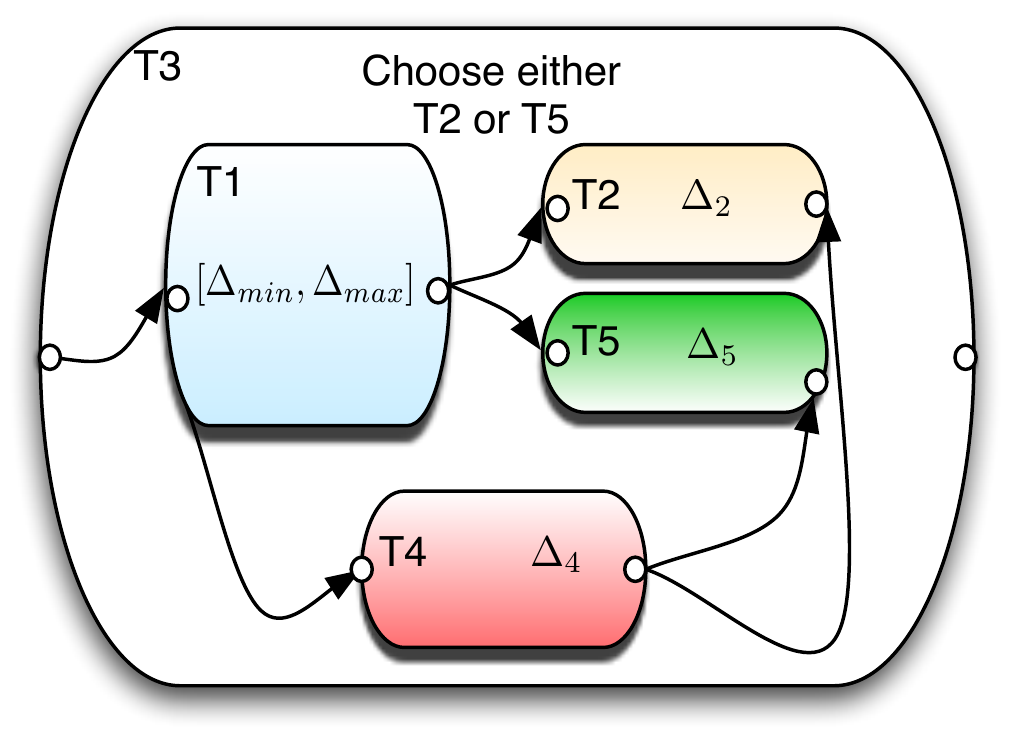}}  
 
      }
    \caption{Limitation of rigid durations.}
    \label{fig:limitations}
  \end{center}
\end{figure}

In some cases (e.g., fig. \ref{fig:basic-example}), we can guarantee that rigid and semi-rigid durations will be respected during the performance of the scenario.
Unfortunately, there is not a generic way to interrupt a rigid TO in a score that contains conditional branching. For instance, figure \ref{fig:limitations} shows a scenario where we cannot preserve the rigid durations of the TOs. $T_1$, $T_4$ and $T_5$ have fixed durations,
but $T_1$ can take different values between $\Delta_{min}$ and $\Delta_{max}$. There is not a way to predict whether $T_2$ or $T_5$ will be chosen after the execution of $T_1$, thus we cannot compute the duration of $T_1$ before the choice.

The problem is that choices do not allow us to predict the duration of the TO's successor; therefore, it is not possible to determinate \textit{a priori} the
duration of the TO. 

\section{Relation to Allombert et al.'s Model}
In this section we show the relation of Allombert \textit{et al.}'s model and ours. First, we briefly present
their model; then, we show the encoding of their model into ours.

\subsection{Allombert et al.'s Model}
Allombert \textit{et al.}'s model is hierarchic: a temporal object (TO) can contain other TOs. A  TO that has children is called a structure. There are two types of structures: \textit{linear structures} have an unidirectional timeline with temporal relations 
inspired on Allen's relations \cite{allen83}; and \textit{logical structures} based on a state-transition model, equivalent a to Finite State Machine. 

The problem of such classification arises when there is a linear structure nested on a logical structure or vice versa: there is not a semantic to handle children of a different type. In what follows, we will focus on representing \textit{linear structures} on our model. Representing \textit{logical structures} is trivial.

We recall some important notations from Allombert \textit{et al.}'s model. A \textit{score} is a tuple $s = \langle T, R \rangle$. A \textit{temporal object} is defined by $t = \langle s,d,\\p,c,N \rangle$, where $s$ is its start date, $d$ is its duration, $p$ is its attached process, $c$ is its local constraint and $N$ are its children. 
A \textit{ temporal relation} is defined by $r = \langle a,t_1,t_2\rangle$, where $a$ is an Allen's relation.
We recall that we do not consider linear temporal relations (e.g., the duration of $A$ is $k$ times the duration of $B$) because they are quantitative temporal relations.

\subsection{Encoding Allombert et al.'s Model}
The function \oen .\cen\  takes a score $s' = \langle T', R' \rangle$ on Allombert \textit{et al.}'s model and returns a score $s = \langle T,R \rangle$ on our model. 
To encode their model, the behavior of each point is to transfer the control to each of its successors and to wait until all its predecessors transfer the control to it. 

A TO $t' = \langle s,d,p,c,N \rangle$ is codified into $t = \langle p_s, p_e, c, d, p, \phi, \{x | x \in $\oen $y$\cen$ \land y \in N\}, \phi\rangle$.  This definition basically codifies a TO and all its children recursively.
We also need to include a Timed Conditional Relation (TCR) $\langle p_s, p_e, \texttt{true}, d, \textit{when}, \\ \textit{wait} \rangle$ for each TO. 

A TR
$r' = \langle a, t_1, t_2 \rangle$ is codified into a TCR $r = \langle p_1, p_2, \texttt{true}, duration, \textit{when}, \textit{wait} \rangle$. Note that the condition is always \texttt{true} and its behavior is \textit{when}, thus the condition is always valid. In addition to the value of the condition, the end behavior \textit{wait} is crucial to define Allen's relations.



\subsubsection{Representing Allen's Relations}
\begin{figure}[h!]
  \begin{center}
    \mbox{
     
{\includegraphics[width=\columnwidth]{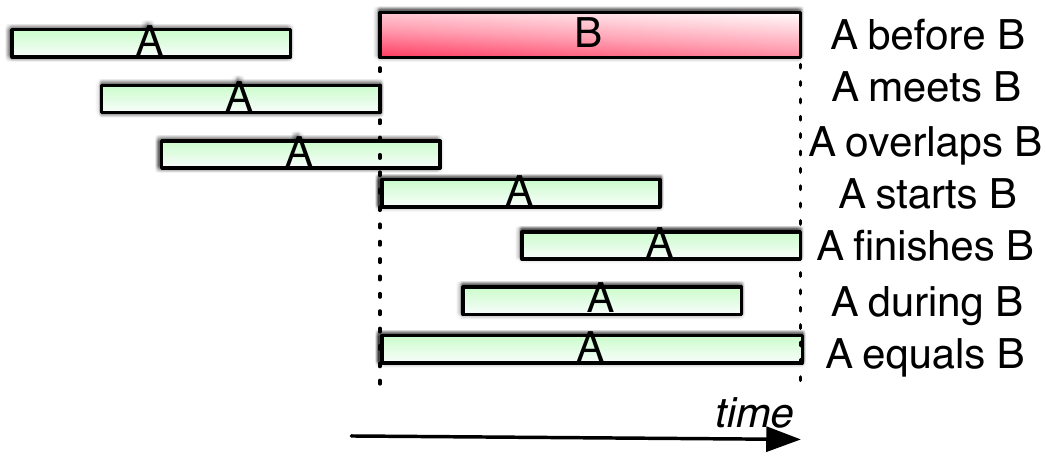}}  
 
      }
    \caption{Allen's relations.}
    \label{fig:allens}
  \end{center}
\end{figure}

Allen's relations represent qualitative TRs between two objects (fig. \ref{fig:allens}).
On Allombert \textit{et al.}'s model, they post constraints during the edition phase to maintain Allen's relations during the execution. For instance, for a relation \textit{B meets A}, they post the constraint $start(A) = start(B) + duration(B)$.

\begin{figure}[h!]
  \begin{center}
    \mbox{
     
{\includegraphics[width=0.5\columnwidth]{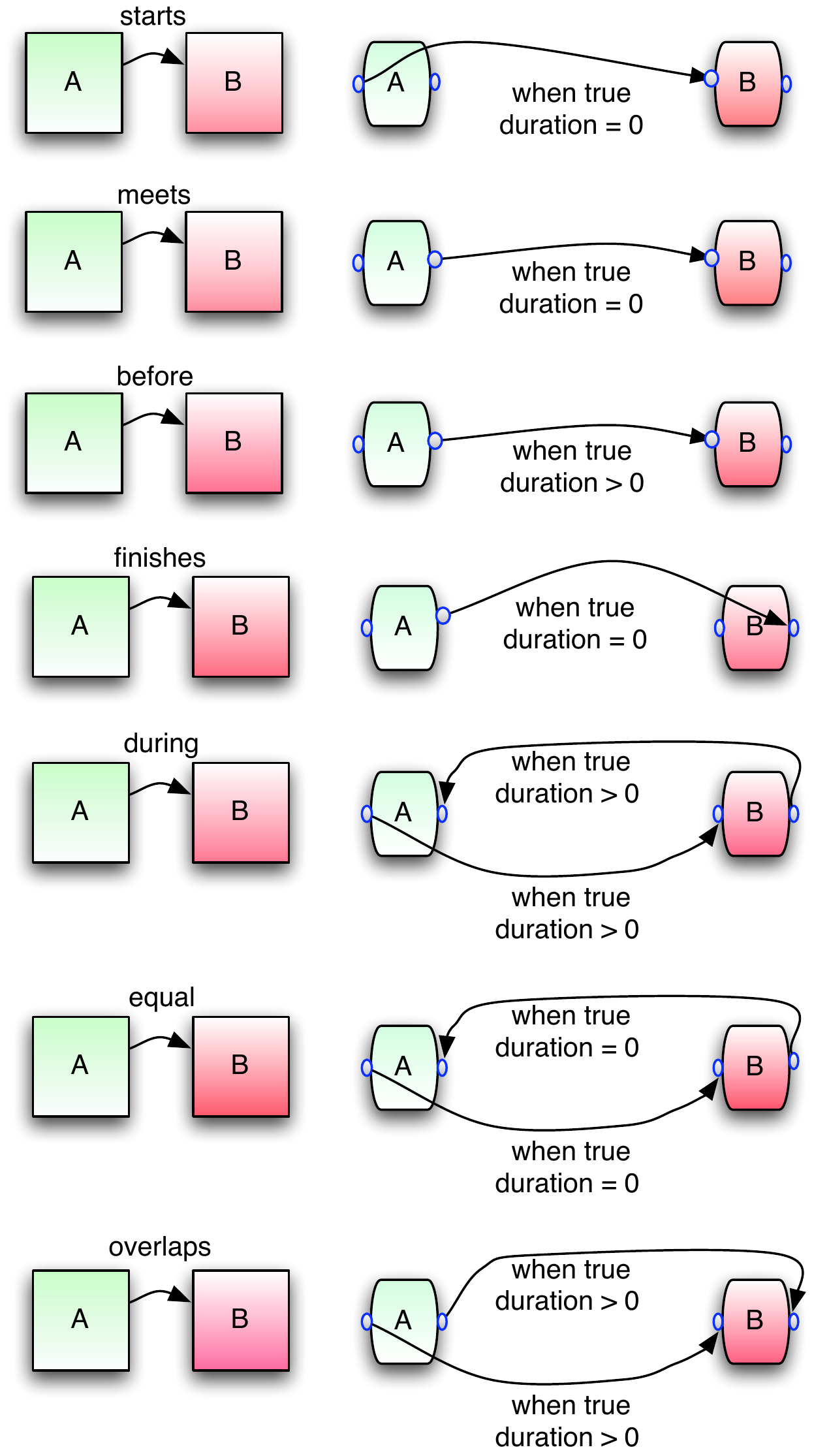}}  
 
      }
    \caption{Representing Allen's relations with Timed Conditional Relations.}
    \label{fig:allens2}
  \end{center}
\end{figure}

We propose  a representation of Allen's relations in terms of our \textit{before} relation. For instance, 
the relation \textit{A meets B} is represented by a relation $\langle e(A),\\ s(B), \texttt{true}, 0, \textit{when}, \textit{wait} \rangle$, where $e$ and $s$ are functions that return the end dates and the start dates of a TO, respectively. Figure \ref{fig:allens2} shows the representation for Allen's relations on our model. 

\section{Study Case: Mariona}
\textit{Mariona} (french acronym for automatic machine with memory, iconography, orinic, narrative and acoustic) is a multimedia installation capable to generate images, analyze
the movements of the users and produce sounds. Its control is described by three temporal objects (TO) that
interact concurrently: \textit{Global}, \textit{Speed}, \textit{Aléatoire} (random). In its current implementation, these TOs are writen on Max/MSP.

\begin{figure}[h!]
  \begin{center}
    \mbox{
     
{\includegraphics[width=0.5\columnwidth]{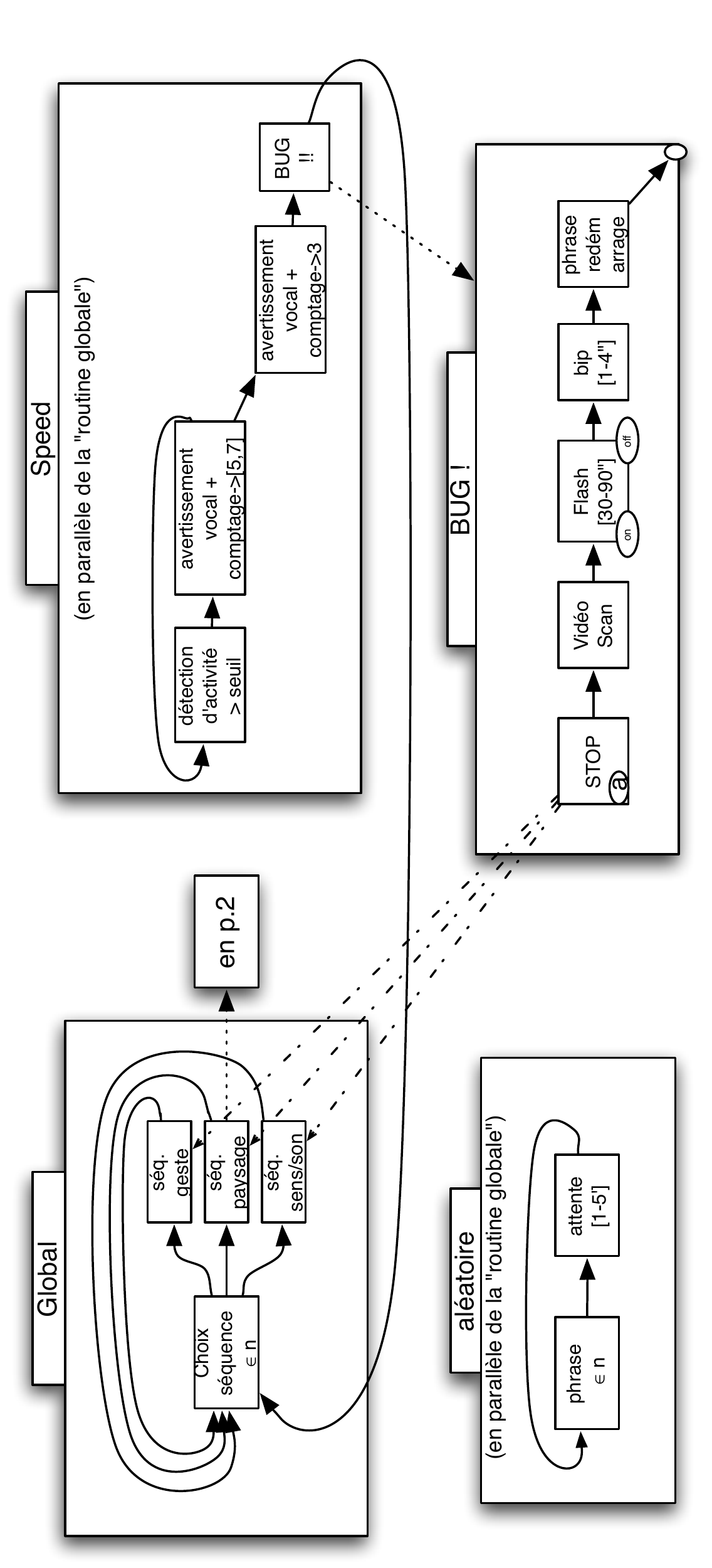}}  
 
      }
    \caption{Mariona: \textit{Global}, \textit{Speed}, \textit{Aléatoire} and \textit{Bug} TOs}
    \label{fig:mariona1}
  \end{center}
\end{figure}

\begin{figure}[h!]
  \begin{center}
    \mbox{
     
{\includegraphics[width=0.5\columnwidth]{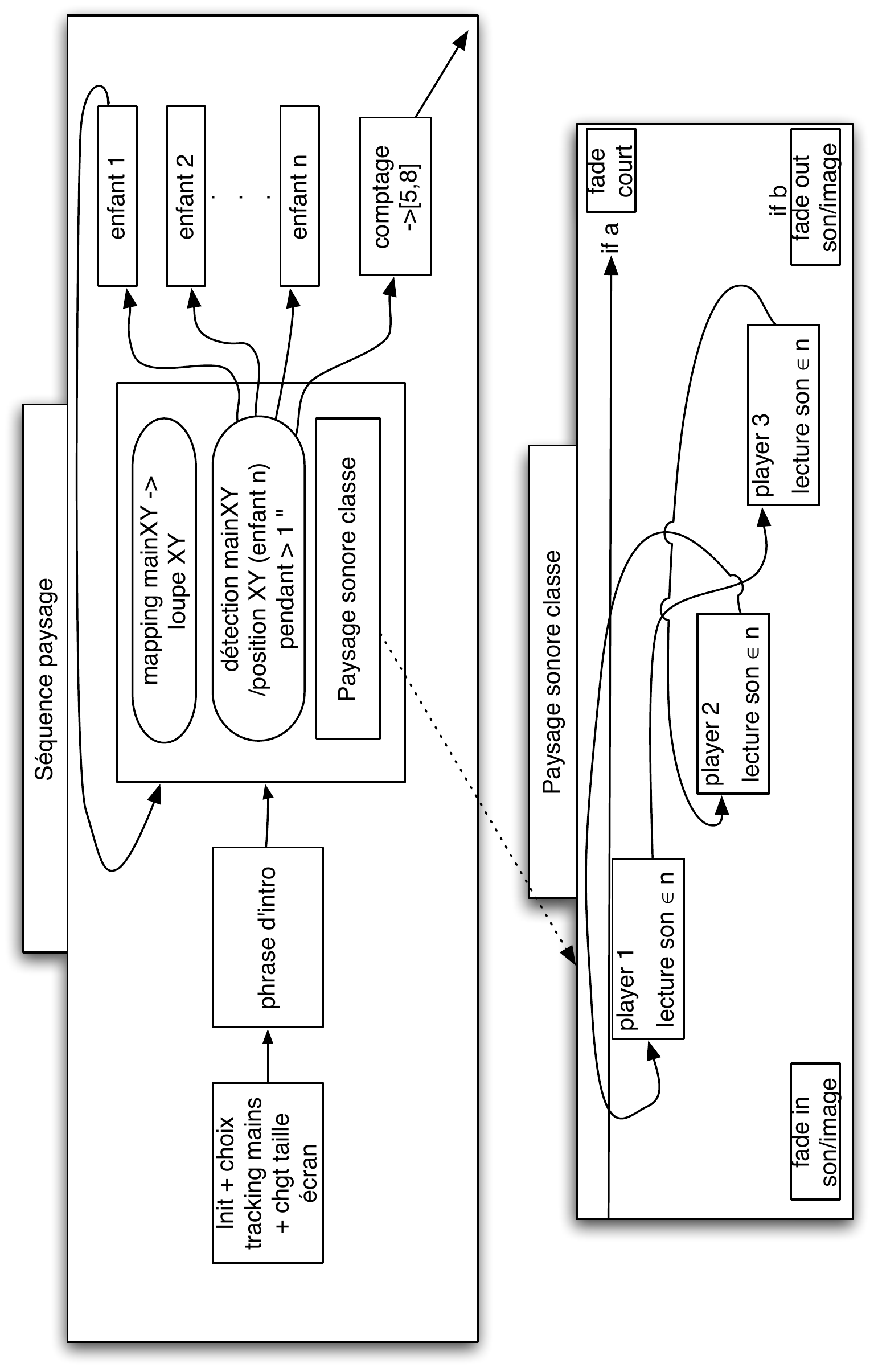}}  
 
      }
    \caption{Mariona: \textit{Séquence paysage} and \textit{Paysage sonore classe} TOs.}
    \label{fig:mariona2}
  \end{center}
\end{figure}

Figure \ref{fig:mariona1} describes the main TOs and \textit{Bug}. Figure \ref{fig:mariona2} describes some TOs contained in \textit{Global}. The notation used in the figures is the one used by Pol Perez.
In Mariona --as we may see in figures \ref{fig:mariona1}  and \ref{fig:mariona2}--,  there is choice, 
trans-hierarchic relations, random durations and loops (finite and infinite).
In what follows, we give some examples about these phenomena and
how we can model them.


\subsection{Choice}
Choice is the main application of conditional branching. It makes it possible to post different relations among TOs
during the edition of the scenario and then choose one during performance. For instance, in the \textit{Global} TO, the system makes a choice among three different TOs: \textit{séq. geste, séq. paysage, séq. sens/son}.

\subsection{Loops and Random Duration}
In the \textit{aléatoire} (random) TO, we may observe loops and a random delay from one to five seconds (\textit{attente [1'-5']}). 
There is also a random duration in the \textit{flash} and \textit{bip} (beep) TOs, which are inside the \textit{Bug} TO.

\subsection{Trans-hierarchical Relations}
Timed Conditional Relations are usually defined between brothers (i.e., two objects that are children of the same TO).
For some scenarios, it is necessary to transfer the control between two TOs with different parents. The conditions for these relations are evaluated in the environment (i.e., the local variables and the local constraint) where it starts. For instance, in Mariona, the \textit{Bug} TO (contained in \textit{Speed}) finishes either \textit{Séq. geste}, \textit{Séq paysage} or \textit{Seq. sens/son} (contained in \textit{Global}). In our model based on
points this does not pose any problem.


     
 





\section{Concluding Remarks}

We presented a new model for Interactive Scores (IS) capable to represent conditional branching together with temporal relations.
The model is based on points.
Temporal objects (TO) and Timed Conditional Relations (TCR) are built upon such points. As far as we know, there are not related work on models for interactive multimedia that support conditional branching besides Allombert \textit{et al.}'s model.

We encoded Allombert \textit{et al.}'s model into our model. We found out that without 
rigid and semi-rigid durations, nor linear temporal relations, it is always possible to transform a temporal relation into  a TCR; although sometimes it is also possible to preserve rigid and semi-rigid durations.
Unfortunately, rigid, semi-rigid and random
durations cannot be always preserved in scenarios with conditional branching.


Whether we can express all the temporal relations presented Allombert \textit{et al.}'s model (e.g., linear temporal relations and rigid durations) into ours, or have to 
choose between a timed conditional bran- ching model and a pure temporal model before defining
a scenario, still remains as an open question.

\subsection{Future Work}
We want
to explore how we can have rigid, semi-rigid durations, and conditional
branching simultaneously. We already determined that it is not possible in the general
case, but we want to know in which cases we can have
such type of durations. We also want to model 
random durations as the ones presented on Mariona.

Another research line is the \textit{waiting behavior} of TOs. For instance, when a TO is supposed to last five seconds, but it lasts ten seconds during performance, what should be the behavior for the aditional five seconds?


Finally, we plan to extend the \texttt{ntcc} model for IS \cite{AADR06} to 
support conditional branching and execute it on \textit{Ntccrt} (a real-time 
capable interpreter for \texttt{ntcc}) \cite{ntccrt}. We also plan to develop
automatic verification tools for Ntccrt in the lines of \cite{fv06}.


\section{Acknowledgements}
We want to thank Antoine Allombert, Matthias Ro- bine and Raphael Marczak for
their valuable comments on our model.



\let\oldbibliography\thebibliography
\renewcommand{\thebibliography}[1]{%
  \oldbibliography{#1}%
  \setlength{\itemsep}{0pt}%
}

\bibliographystyle{abbrv}
{\scriptsize
\bibliography{mybib}
}




\end{document}